\documentclass[journal]{IEEEtran} % double column single space
\usepackage{booktabs}
\usepackage[dvipsnames]{xcolor}
\usepackage{cite} 
\usepackage[pdftex]{graphicx}
\graphicspath{ {./image/} }
\usepackage[cmex10]{amsmath} 
\usepackage{dsfont}% to use in place of bbm; \mathds{}
\usepackage{array} 
\ifCLASSOPTIONcompsoc
\usepackage[caption=false,font=normalsize,labelfont=sf,textfont=sf]{subfig}
\else
\usepackage[caption=false,font=footnotesize]{subfig}
\usepackage{diagbox}
\usepackage{booktabs}
\usepackage{ wasysym }
\usepackage[T1]{fontenc} 
\fi
\allowdisplaybreaks
\usepackage{color}
\usepackage{float}
\usepackage{fixmath, nicefrac}
\usepackage{stfloats}
\usepackage{arydshln} 
\usepackage[]{changes}
\usepackage{url}
\usepackage{tabularx}
\usepackage{textcomp} % copyright symbol
\usepackage{adjustbox} % scaling figure size
\usepackage{textcase}
\usepackage{graphicx}
\usepackage{enumerate}
\usepackage{footnote}
\usepackage{xparse}
\usepackage{amssymb}
\usepackage{amsmath}
\usepackage{amsthm}
\usepackage{amsfonts}
\RequirePackage{xspace}
\RequirePackage{graphics}
\RequirePackage{textcomp}
\usepackage{keyval}
\usepackage{xspace}
\usepackage{mathrsfs}
\usepackage{paralist}
\usepackage{lipsum}
\usepackage{listings}
\usepackage{multirow}
\usepackage{array}
\usepackage{pifont} % check
\usepackage{stmaryrd}
\usepackage{textcase}
\usepackage{makecell}
\usepackage{stackengine}%
\usepackage{threeparttable}
\usepackage{mathtools}
\usepackage[T1]{fontenc}
\usepackage{bm}
\usepackage{amsmath}

\newcommand{\seclabel}[1]{\label{sec:#1}}

\renewcommand{\eqref}[1]{~(\ref{eq:#1})}

\begin{document}

\title{Interoperability of Electric Drive Models using HELICS}

\author{Ajay~Pratap~Yadav,~\IEEEmembership{Member,~IEEE}%
	\thanks{Ajay Pratap Yadav is with Oak Ridge National Laboratory, TN, USA. 
		(e-mail: yadavap@ornl.gov).}
}

 \maketitle

\begin{abstract}
Accurate modeling and simulation are essential for the effective design, testing, and evaluation of electric machine systems. However, existing models often face interoperability challenges due to differences in programming languages (e.g., C, MATLAB, Python) and the separation of components such as inverters and controllers across diverse environments. These challenges are amplified by the growing use of advanced simulation platforms like Hardware-in-the-Loop (HIL) and Controller-HIL, which require repeated adaptations for compatibility. This paper presents a HELICS-based co-simulation framework that enables seamless coordination among heterogeneous tools and models, providing a unified platform for integrating and testing electric drive models regardless of their origin. The approach is demonstrated through the co-simulation of an inverter-fed permanent magnet synchronous machine (PMSM) drive under speed control, showcasing reduced development time, flexible reuse of existing models, and efficient integration into both software and HIL environments—offering a scalable, modular solution for collaborative and repeatable electric drive system testing and development.  
\end{abstract}

\footnote{This manuscript has been authored by UT-Battelle, LLC, under contract DE-AC05-00OR22725 with the US Department of Energy (DOE). The US government retains and the publisher, by accepting the article for publication, acknowledges that the US government retains a nonexclusive, paid-up, irrevocable, worldwide license to publish or reproduce the published form of this manuscript, or allow others to do so, for US government purposes. DOE will provide public access to these results of federally sponsored research in accordance with the DOE Public Access Plan (http://energy.gov/downloads/doe-public-access-plan).}

\section{Introduction}
The Hierarchical Engine for Large-scale Infrastructure Co-Simulation (HELICS) is a co-simulation framework designed to facilitate seamless integration of heterogeneous simulation models across multiple domains~\cite{8064542}. It enables communication and synchronization among models developed in different programming languages and environments, and has been effectively applied to large-scale power system co-simulations~\cite{wang2021transmission}. In this paper, we explore how HELICS can be utilized to build an interoperable framework for testing electric drive models.

Modeling and testing of electric machines is a critical component of their design and development. These models can be built using a range of physics-based approaches, such as finite-element methods (FEM) and magnetic equivalent circuit (MEC) models, or through lumped-parameter representations like phase-domain (PD), direct–quadrature (qd0), and voltage-behind-reactance (VBR) models. In practice, engineering constraints and user preferences often lead to different components being implemented in various programming languages or simulation environments.

HELICS addresses these interoperability challenges by enabling co-simulation of heterogeneous components, thus promoting the reuse of legacy models and supporting rapid, modular development. Built with scalability in mind, HELICS is applicable to both small- and large-scale systems and facilitates the integration of high-fidelity machine and control models for comprehensive testing. The advantage of this approach is that each drive component can remain an independently executed federate with its own implementation language, solver, dependencies, and internal time step. This is useful when controller code, machine models, inverter models, or external test interfaces are developed by different teams and should be connected through a stable signal contract rather than rewritten into a single tool. HELICS was designed for large infrastructure co-simulation, so it is also useful when a drive model must interact with grid, controller, communication, or hardware interfaces.

\begin{figure}
	\centering
	\includegraphics[width=\linewidth]{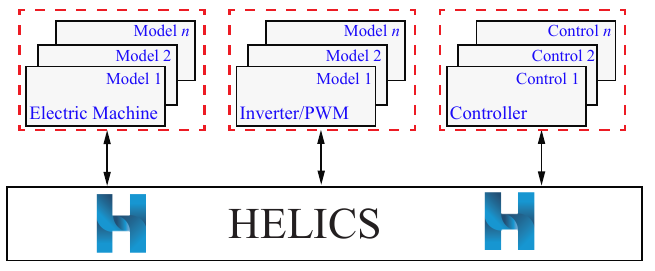}
	\caption{Modular HELICS arrangement for electric-drive studies. Controller or machine implementations can be exchanged without changing the interface.}
	\label{fig_HELICS}
\end{figure}

In this paper, we present a HELICS-based co-simulation framework for evaluating electric drive systems as shown in  Figure \ref{fig_HELICS}. This modular setup enables flexible model development and integration, and its utility is demonstrated through simulation results obtained from conventional machine drive scenarios.

\section{Electric Drive Description}
\begin{figure}
	\centering
	\includegraphics[width=\linewidth]{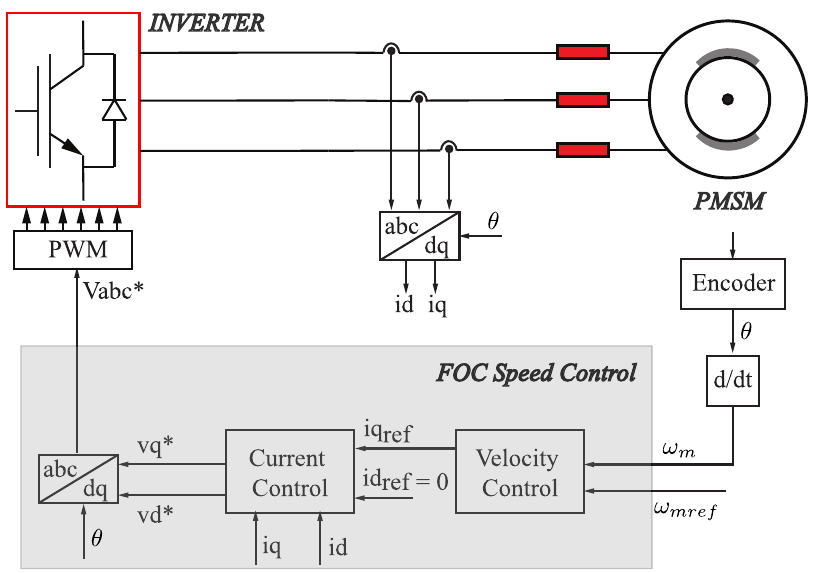}
	\caption{PMSM speed control using the FOC strategy.}
	\label{fig_FOC}
\end{figure}

\seclabel{sec_2}
Figure~\ref{fig_FOC} shows an inverter-fed PMSM controlled via field-oriented control (FOC), featuring two PI controllers for speed and current regulation.

The PMSM dynamics are modeled as~\cite{9804715}:
\begin{align}
	&	\frac{d i_{d}}{dt} = \frac{1}{L_d} \big( v_{d} - R_s i_{d} + P_p \omega_m L_q i_{q} \big), \label{eq_PMSM_voltage1} \\
	&	\frac{d i_{q}}{dt} = \frac{1}{L_q} \big( v_{q} - R_s i_{q} - P_p \omega_m L_d i_{d} - P_p \omega_m \lambda_m \big), \label{eq_PMSM_voltage2} \\
	&	\frac{d \omega_m}{dt} \!=\!\! \frac{1}{J_m} \!\left( \!\frac{3 P_p}{2} \big( \lambda_m i_{q} \!+\! (L_d \!-\! L_q) i_{q} i_{d} \big) \!-\! d_m \omega_m \!-\! T_L \! \right)\!, \label{eq_PMSM_3} \\
	&	\frac{d \theta_m}{dt} = \omega_m. \label{eq_PMSM_4}
\end{align}
Here, $i_q$ and $i_d$ are the $q$- and $d$-axis stator currents; $\omega_m$, the rotor speed; and $\theta_m$, the mechanical rotor angle. Parameters $R_s$, $L_d$, and $L_q$ denote stator resistance and $d$/$q$-axis inductances. $P_p$ is the pole pair count, and $\lambda_m$ is the permanent magnet flux linkage.

Park and inverse Park transforms relate the $abc$ and $dq$ frames. For a general signal $f$:
\begin{align}
	&[f_q, f_d]^\top = \mathbf{K_s}(\theta_r) \, [f_a, f_b, f_c]^\top, \\
	&[f_a, f_b, f_c]^\top = \mathbf{K_s}^{-1}(\theta_r) \, [f_q, f_d]^\top,
\end{align}
with $\theta_r = P_p \theta_m$ being the electrical rotor angle~\cite{erickson2007fundamentals}.

The inverter output is modeled as:
\begin{equation}
	v_{f=a,b,c}(t) =
	\begin{cases}
		V_{dc}, & v^*_{f=a,b,c}(t) \ge z(t), \\
		0,      & v^*_{f=a,b,c}(t) < z(t),
	\end{cases}
	\label{eq:inv_phasec_combine}
\end{equation}
where $f$ denotes phases $a$, $b$, $c$, and $z(t)$ is a unit-amplitude triangular carrier at the switching frequency. The reference voltages $v^*_{a,b,c}$ are balanced three-phase cosines, with $\theta_s$ as the phase angle of the reference voltage vector.
\subsubsection*{Control Equations}
FOC employs PI controllers for speed and current control:
\begin{align}
&	\frac{d \phi}{dt} = \omega_{mref} - \omega_m \\
&	i_{qref} = K_{PV} (\omega_{mref} - \omega_m) + K_{IV} \phi \\
&	i_{dref} = 0 \\
&	\frac{d \psi_q}{dt} =  i_{qref} - i_q \\
&	\frac{d \psi_d}{dt} =  i_{dref} - i_d \\
&	v_q^* = K_{PC}(i_{qref} - i_q ) + K_{IC} \psi_q \\
&	v_d^* = K_{PC}(i_{dref} - i_d ) + K_{IC} \psi_d \label{eq:74}
\end{align}
The controller states $\phi$, $\psi_d$, and $\psi_q$ generate the reference voltages $v_q^*$ and $v_d^*$ for the PWM. The parameters $K_{PV}$, $K_{IV}$, $K_{PC}$, and $K_{IC}$ represent the PI controller gains.

\section{HELICS Co-Simulation Setup}
\subsection{Overview}
HELICS is a co-simulation framework that enables multiple independent simulators to operate within a coordinated simulation \cite{10424422}. It provides two core functions: synchronizing simulation time across federates and routing data between them. This allows domain-specific simulators to remain separate while exchanging information over a shared simulation timeline.

A HELICS co-simulation includes two main components: \textit{federates}, which are individual simulators, models, or applications that exchange values, messages, or both; and the \textit{broker/core}, which manages time synchronization and data routing. Federates connect to a \textit{core}, which communicates with a \textit{broker}; larger simulations can use hierarchical brokers for scalability.
\subsection{Data exchange and Time}
For data exchange, HELICS provides two main interfaces. The \textit{publication/subscription} interface supports typed, unit-aware data streams such as voltage measurements, controller setpoints, sensor outputs, and other time-series variables. The \textit{endpoint/message} interface supports packet-style communication through named endpoints and is useful when information is better represented as a discrete message. A federate can use both interfaces within the same model, enabling flexible simulator coupling.

Time advancement in HELICS is controlled through explicit time requests. Each federate requests permission to advance using \texttt{requestTime}, and HELICS grants time only when the ordering of data exchange is preserved. For iterative convergence, \texttt{requestTimeIterative} allows coupled simulators to exchange values repeatedly at the same simulation time before advancing.

HELICS supports several programming environments, including C/C++, Python, MATLAB, Octave, Java, Julia, and Nim. This flexibility allows it to connect heterogeneous simulation tools within a single workflow. For example, HELICS has been used to facilitate data exchange with real-time simulation platforms such as OpalRT~\cite{10424422}.
\section{Results}
We consider two cases, each using the same HELICS interface. The plant federate (machine model) publishes $i_q$, $i_d$, $\omega_m$, and $\theta_m$ and subscribes to $v_q^*$ and $v_d^*$; the controller federate performs the inverse exchange. HELICS grants time in $100~\mu$s increments, while PWM-based plants integrate internally with a $10~\mu$s step. The drive starts from $i_q=2$ A and zero remaining states, follows an $80/120/100$ rad/s speed command with steps at $0.65$ s and $1.35$ s, and receives a $5$ N.m load at $1.0$ s.

\begin{figure}[t]
	\centering
	\includegraphics[width=\linewidth]{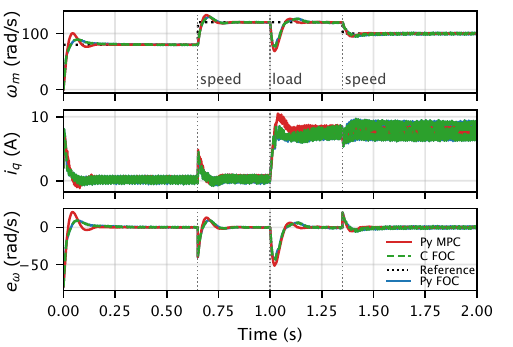}
	\caption{Python PMSM/inverter plant coupled to Python FOC, Python one-step MPC, and C FOC controllers through the same HELICS interface.}
	\label{fig_controller_comparison}
\end{figure}

Figure~\ref{fig_controller_comparison} compares three controllers on the same Python PMSM/inverter plant. Python FOC and C FOC give the same response within plotting resolution, with speed RMSE $6.42$ rad/s, final speed error $0.62$ rad/s, and peak $|i_q|=9.70$ A. The one-step MPC is inserted through the same interface and remains comparable, with speed RMSE $7.32$ rad/s, final speed error $0.45$ rad/s, and peak $|i_q|=10.52$ A.

\begin{figure}[t]
	\centering
	\subfloat[Nominal PWM PMSM.\label{fig_machine_nominal_pwm}]{%
		\includegraphics[width=\linewidth]{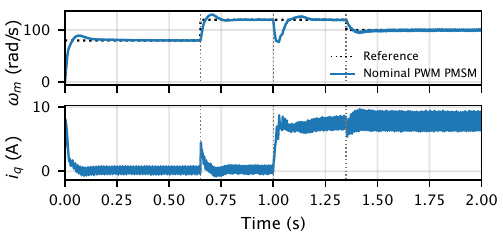}}
	\\[-0.3em]
	\subfloat[Parameter-varied PWM PMSM.\label{fig_machine_parameter_pwm}]{%
		\includegraphics[width=\linewidth]{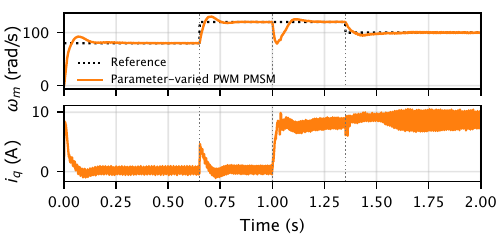}}
	\\[-0.3em]
	\subfloat[MATLAB/Octave-compatible averaged PMSM.\label{fig_machine_averaged_pmsm}]{%
		\includegraphics[width=\linewidth]{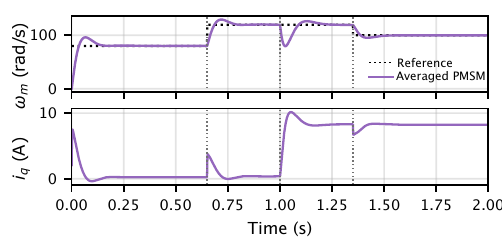}}
	\caption{Three different machine models coupled through the same HELICS interface to the same C FOC controller}
	\label{fig_machine_cases}
\end{figure}

The second study holds the C FOC controller fixed and exchanges only the plant federate. The nominal PWM PMSM in Fig.~\ref{fig_machine_cases}(a) uses inverter switching; the parameter-varied PWM PMSM in Fig.~\ref{fig_machine_cases}(b) uses the same switched-inverter structure with different machine constants; the averaged PMSM in Fig.~\ref{fig_machine_cases}(c) applies the normalized command directly as averaged $dq$ voltage.
The exchanged plants have similar speed-tracking metrics under the same C controller: RMSE/final-error pairs are $6.42/0.62$ rad/s for the nominal PWM model, $6.56/0.56$ rad/s for the parameter-varied PWM model, and $6.94/0.00$ rad/s for the averaged PMSM. Current traces are shown per machine in Fig.~\ref{fig_machine_cases}, but are not overlaid because the models use different parameterizations.

\section{Conclusion}
This letter demonstrates a HELICS-based interoperability workflow for electric drives. The results show that equivalent controllers in different languages produce equivalent plant responses, while controller and plant implementations can be exchanged independently. The current study is intentionally limited to software co-simulation; future work will extend the same interface to HIL targets and larger drive systems.

\appendices
\section{System and Simulation Parameters}
\label{app:parameters}
The dc link is a stiff $V_{dc}=200$ V source with no modeled dc-link capacitor, ac filter, or semiconductor loss. PWM cases use an ideal two-level inverter at $f_{sw}=10$ kHz, while the averaged case applies $v_q=0.5V_{dc}v_q^*$ and $v_d=0.5V_{dc}v_d^*$. 
The nominal PWM machine uses $R_s=0.636~\Omega$, $P_p=5$, $L_d/L_q=20/12$ mH, $\lambda_m=0.088$ Wb, and $J_m=1.0{\times}10^{-3}$ kg.m$^2$. The parameter-varied PWM case changes the main constants to $R_s=0.820~\Omega$, $L_d/L_q=17/9$ mH, $\lambda_m=0.080$ Wb, and $J_m=1.25{\times}10^{-3}$ kg.m$^2$. The averaged PMSM uses $R_s=0.720~\Omega$, $P_p=4$, $L_d=L_q=11$ mH, $\lambda_m=0.105$ Wb, and $J_m=1.45{\times}10^{-3}$ kg.m$^2$. The Python and C FOC controllers use $K_{PV}=0.1$, $K_{IV}=5.0$, $K_{PC}=0.1$, and $K_{IC}=1.0$ with $i_{dref}=0$ and $i_{qref}$ limited to $\pm18$ A; the one-step MPC uses a $600~\mu$s horizon and a $17\times17$ voltage-command grid.

\bibliographystyle{IEEEtran}
\bibliography{mode_references.bib}

\end{document}